\documentclass[12pt, preprintnumbers, amsmath, amssymb]{revtex4-1}
\usepackage[draft]{hyperref}
\usepackage{amsmath}
\usepackage{tabularx,amsmath,caption,setspace,float,graphicx,dcolumn,bm,color,bbding,caption,footnote,array}
\begin{document}
\title{Wide-span and tunable-wavelength photon pairs around 1.55 $\mu$m\\ from a silicon nanophotonic chip}

\author{Ranjeet Kumar}\email{rkumar@ucsd.edu}\affiliation{Department of Electrical \& Computer Engineering, University of California, San Diego, La Jolla, California 92093}
\author{Jun Rong Ong}\affiliation{Department of Electrical \& Computer Engineering, University of California, San Diego, La Jolla, California 92093}
\author{John Recchio}\affiliation{Department of Electrical \& Computer Engineering, University of California, San Diego, La Jolla, California 92093}
\author{Kartik Srinivasan}\affiliation{Centre for Nanoscale Science \& Technology, National Institute of Standards \& Technology,
Gaithersburg, Maryland 20899}
\author{Shayan Mookherjea}\affiliation{Department of Electrical \& Computer Engineering, University of California, San Diego, La Jolla, California 92093}
\begin{abstract}
Using a compact optically-pumped silicon nanophotonic chip  consisting of coupled silicon microrings, we generate photon pairs in multiple pairs of wavelengths around 1.55 $\mu$m. The wavelengths are tunable over several nanometers, demonstrating the capability to generate wavelength division multiplexed photon pairs at freely-chosen telecommunications-band wavelengths. 
\end{abstract}
\maketitle

Correlated photon pairs have traditionally been generated using bulk crystals and in optical fibers; however, SOI (silicon-on-insulator) based devices may be advantageous ~\cite{sharping06,harada08} because of the high refractive index contrast between the silicon core and the silica cladding, which increases the mode confinement, enhances nonlinearity, and thus reduces the device size. Photon pairs are generated using spontaneous four-wave mixing, where two pump photons each of frequency $\omega_{p}$ from a single pump beam generate a pair of photons at the Stokes  ($\omega_{s}$) and anti-Stokes ($\omega_{i}$) frequencies. Periodically patterned silicon nanophotonic waveguides can have an effective {K}err nonlinearity coefficient$\gamma_\textrm{eff} \approx 4,000 \ \text{W}^{-1}\text{m}^{-1}$ to $10,000\ \text{W}^{-1}\text{m}^{-1}$ that is 5 to 6 orders of magnitude larger than that of Highly Non-Linear Fiber (HNLF) around 1.5 $\mu$m ~\cite{melloni11, ong11, mastuda11}, allowing for pair generation using only milliwatts of optical pump powers, in a regime where two-photon absorption and free-carrier generation losses may be small. Silicon microrings and microring arrays have been used in several recent demonstrations of photon pair generation and heralded single photon generation ~\cite{davanco12,clemmen09,engin12,jiang12,azzini12,sharping06,harada08}.


Compared our previous report~\cite{davanco12}, here we have improved (lowered) the propagation loss, and designed a more compact structure with fewer cascaded rings which allowed us to demonstrate a significantly higher coincidence-to-accidental ratio (CAR). Furthermore, we report two other advancements: (1) We demonstrate generation of photon pairs whose wavelengths can be tuned, using electrical current-driven thermo-optic change of the refractive index. (2) We demonstrate experimentally that a device based on silicon microrings can generate multiple lines of Stokes and anti-Stokes photon pairs coupled into the same output waveguide.  

\begin{figure}[h]
\centerline{\includegraphics[scale=0.91,trim = 0cm 0cm 0cm 0.5cm]{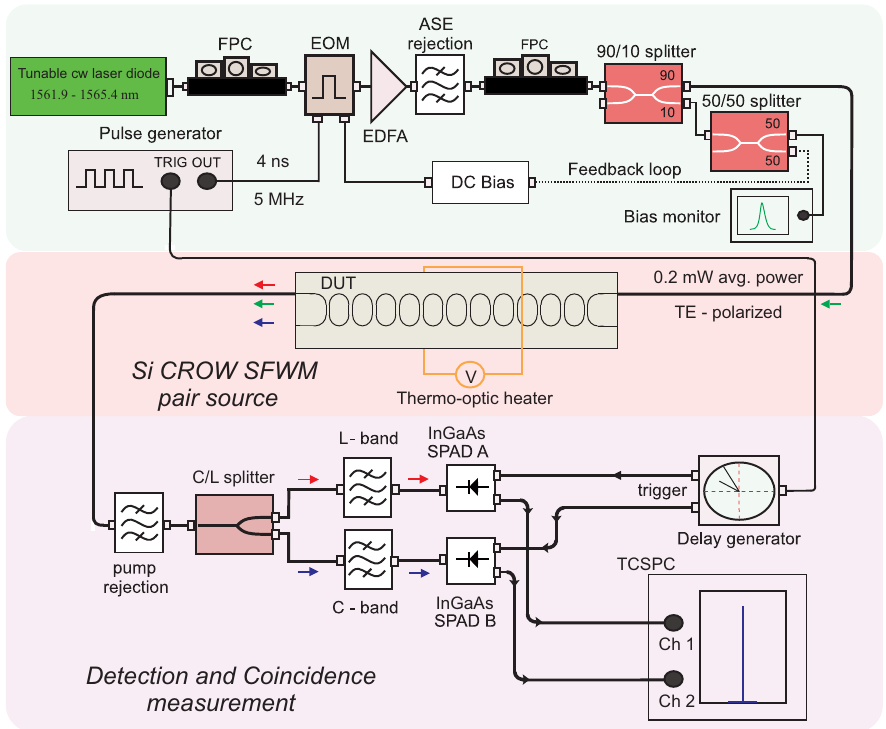}}
\caption{Photon pair generation using diode-pumped spontaneous four-wave mixing (SFWM). EDFA = Erbium doped fiber amplifier, FPC = Fiber polarization controller, EOM = Electro-optic modulator, ASE = Amplified spontaneous emission, DUT = Device under test, SPAD = Single photon avalanche photo diode, TCSPC = Time-correlated single photon counting.\label{setupp}}
\end{figure}
\begin{figure}[h]
\centerline{\includegraphics[scale=0.40, trim = 0cm 3cm 0cm 1.5cm]{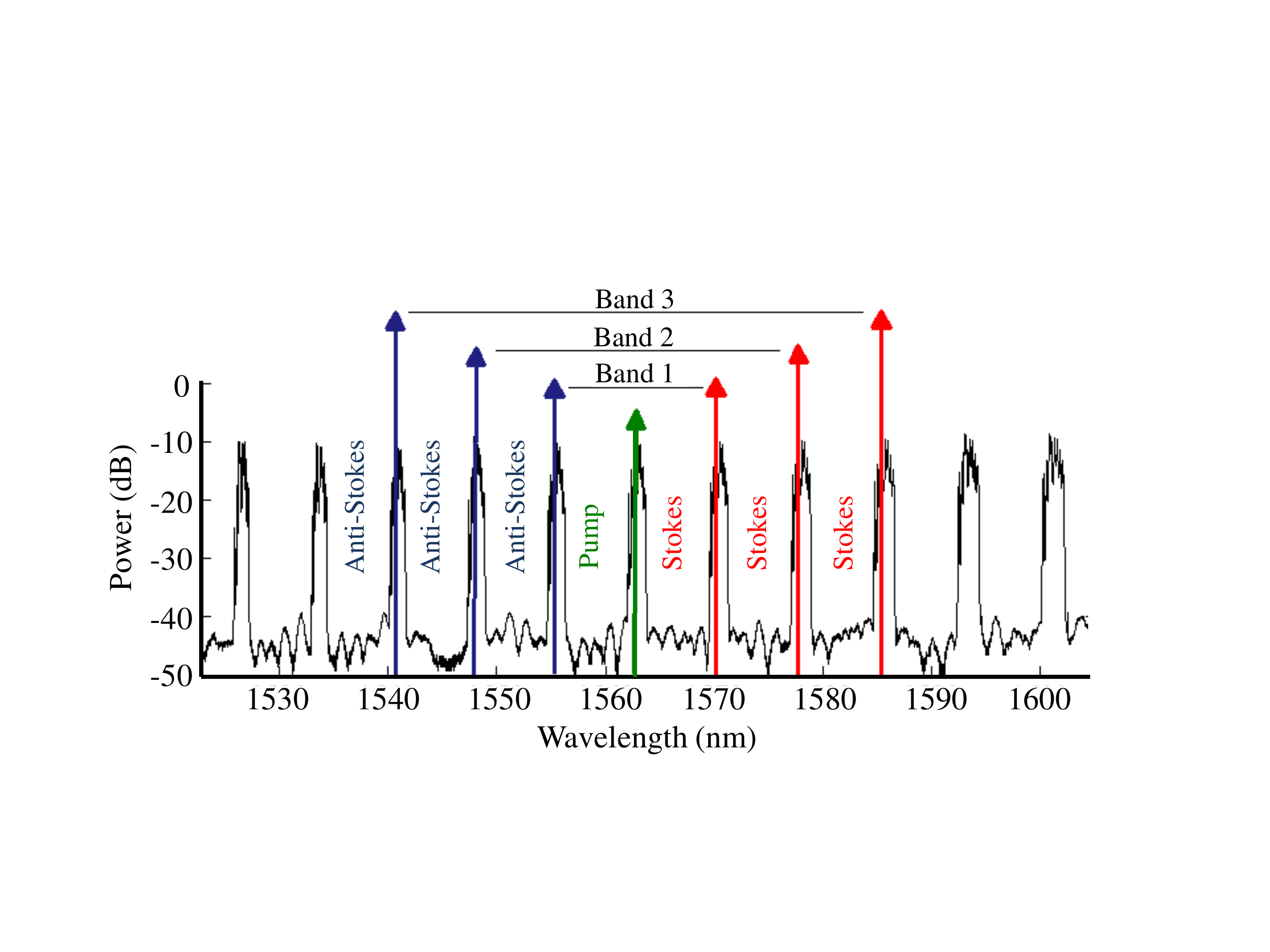}}
\caption{A classical transmission spectrum of the 11-ring coupled-microring device at room temperature (296.35~K) showing a series of passbands and stopbands in a clean, single-mode family, with high passband-stopband contrast. Using a single pump, multiple sets of photon pairs were generated, as indicated. Three such pairs were sequentially measured, as reported in Table.~\ref{table1}.}
\label{trans}
\end{figure}

The experimental setup is shown in Fig.~\ref{setupp}. Photon pairs are generated in a coupled-resonator optical waveguide (CROW) consisting of 11 cascaded microrings, cumulatively spanning a distance of 0.23~mm on the silicon chip. The transmission spectrum of the CROW is shown in Fig.~\ref{trans}, with several passbands of about 1.75~nm in spectral width separated by the 7 nm free spectral range of the constituent microrings. The devices were fabricated using CMOS-compatible processes on SOI wafers with 220 nm silicon layer height, and singulated into chips for testing using edge-coupled waveguide-to-fiber tapers. The insertion loss of each fiber-to-waveguide coupler was estimated as 4.3~dB, based on calibration measurements on separate test sites. Light was transmitted through the CROW in a disorder-tolerant slow light regime, with slowing factor $S = n_{g, CROW}/n_{g,WG}$ between 32 and 42, depending on the wavelength ($n_{g, CROW}$ is the group index of the CROW; $n_{g, WG}$ is the group index of a conventional waveguide), which is about 3x higher than in our earlier report~\cite{davanco12}. The propagation loss was about  0.13~dB per ring including slow light enhancement of the loss, i.e.,~an insertion loss of about 1.4~dB for the microring section (about 6~dB improvement from Ref.~\cite{davanco12}). Compared to a conventional waveguide, the rate of spontaneous pair generation can be enhanced by a factor of $S^4$ in an ideal CROW. Compared to a single microring, the CROW can offer an $N^2$ enhancement in the generation rate, or, alternatively, wider passbands for ease of experimental alignment, and increased thermal stability~\cite{ong13}. 


Using the setup shown in Fig.~\ref{setupp}, we excited a spontaneous four-wave mixing (SFWM) process ~\cite{clemmen09,engin12,jiang12,azzini12,davanco12,sharping06,harada08}, generating polarization-degenerate and frequency non-degenerate photon pairs. In view of the ripples in the transmission passbands (shown in Fig.~\ref{trans}), the pump wavelength was initially chosen by optimizing the classical four-wave mixing conversion of a weak signal beam at the Stokes wavelength to the anti-Stokes wavelength. TE-polarized light at 1562.6 nm with 4 ns pulse width, 5 MHz repetition rate, 200 $\mathrm{\mu W}$ average power was used for the measurements in Table~\ref{table1}, where we report on the three nearest bands relative to a fixed pump wavelength, as shown in Fig.~\ref{trans}. The generated Stokes and anti-Stokes photons were separated using a tunable set of narrowband filters with full-width at half-maximum (FWHM) of 0.6 nm for the C-band and 1.0 nm for the  L-band, with insertion loss of 6.2 dB in each case, and passband-to-stopband contrast exceeding 150~dB. Photons were detected using InGaAs single photon avalanche detectors (SPADs) using a reverse bias voltage of 3.0 V, resulting in an estimated quantum efficiency of 10~\%. The SPADs were electrically gated with a window of 4 ns, synchronously with the optical pump repetition rate. The averaged detector dark count rates in the two channels were measured to be 94 Hz and 195 Hz. The measured average singles count rates were 98 kHz for the anti-Stokes photons in the C-band and 76 kHz for the Stokes photons in the L-band. Table~\ref{table1} reports the coincidence-to-accidental ratio (CAR), which is a metric calculated as described in Refs.~\cite{davanco12,clemmen09,engin12,jiang12,azzini12,sharping06,harada08} among others, that depends on various parameters such as the pair generation rate, detector dark counts, detector gating window, multi-pair generation probability and optical propagation losses in the measurement pathway. Generally, a value of CAR $\textgreater$ 10, as shown in Table~\ref{table1}, shows adequate utility of the device as a quantum light source \cite{sharping06,harada08}. These results show a 6x improvement in CAR compared to Ref. ~\cite{davanco12}. 

\begin{figure*}[h]
\centerline{\includegraphics[scale =0.88,trim = 0cm 0.5cm 0cm 0cm]{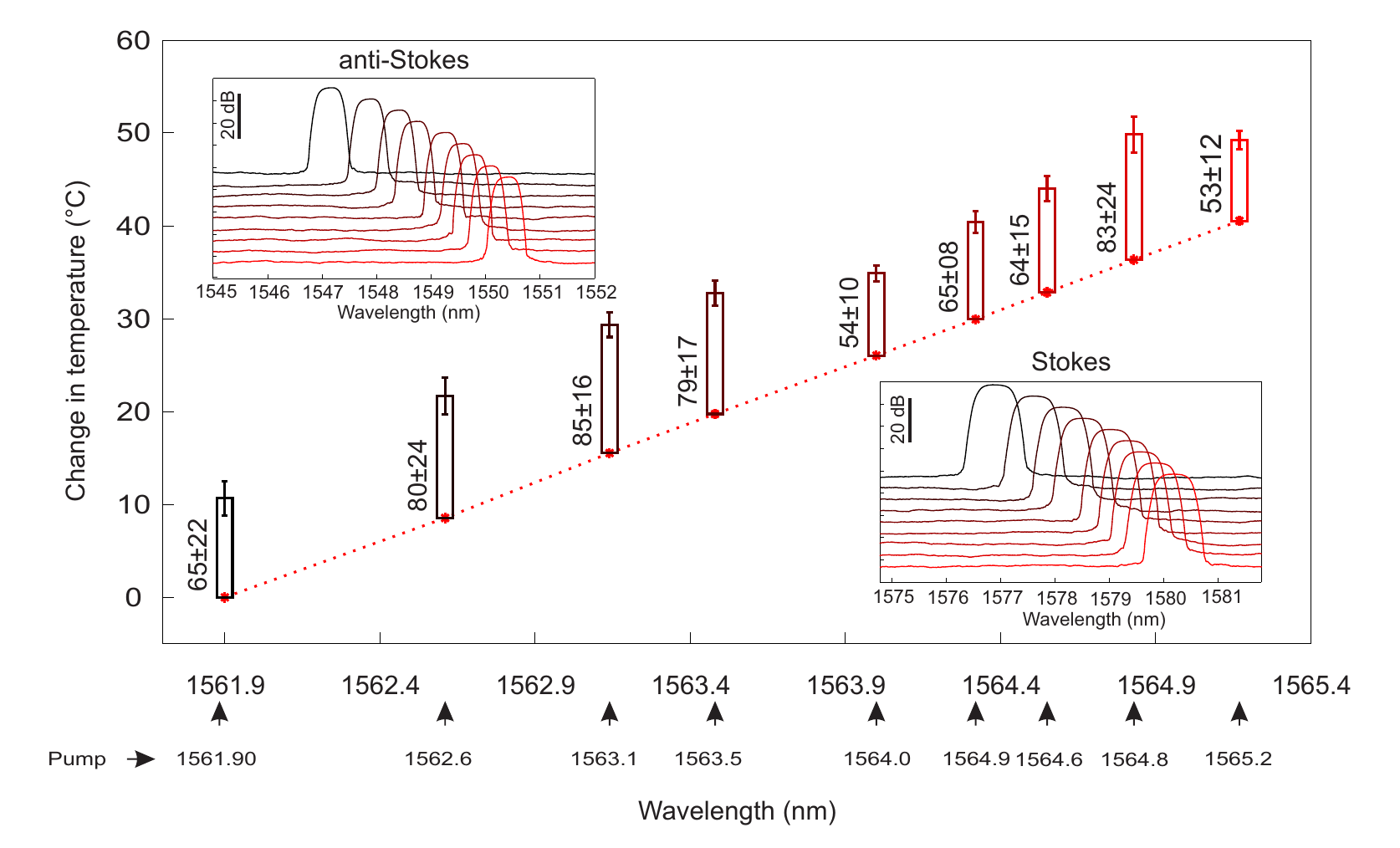}}
\caption{Changing the temperature of the chip changes the wavelengths of the Stokes and anti-Stokes photon pairs, and requires a linear tuning of the input pump wavelength. The height of the bars is proportional to the measured coincidence-to-accidental ratio (CAR) between the photon pairs. Error bars in the CAR values are determined as in Table~\ref{table1}. \emph{Insets}: Peak pair generation wavelengths for anti-Stokes (top - left) and Stokes (bottom - right) as shown here by tuned filtering channels, in front of the wavelength insensitive SPADs.}
\label{tunecar}
\end{figure*}

When the chip was thermally heated or cooled using a thermo-electric module, the spectrum shifted cleanly, and without loss of contrast. As shown by the insets to Fig.~3, both the anti-Stokes and the Stokes wavelengths at which pairs were generated shifted smoothly to longer wavelengths upon heating, without significant change in their spectral separation ($\lambda_\text{Stokes}-\lambda_\text{anti-Stokes}$). Thus, in this device, photon pairs can be generated at any wavelength, e.g., aligned to a grid of lambdas prescribed by a fiber network, if the free spectral range of the constituent microrings is also compliant with this grid. To demonstrate this tunability, we changed the temperature of the chip, using an electrical current driven thermo-optic effect, and tuned the input pump wavelength and the output filters, in accordance with the predictions of energy and momentum conservation. In Table~\ref{table2}, we report the pair generation rate for Band 2 (indicated in Fig.~2) at several different temperatures; Bands 1 and 3 have similar temperature tuning behavior. In Fig.~\ref{tunecar}, we show that the wavelength shift was linear with temperature, and a high CAR value was maintained while tuning. This shows that such a device may be useful for wavelength-division multiplexing (WDM) of silicon photon pair sources. 

In summary, we demonstrated photon pair generation from a comb of frequencies produced by a coupled-microring resonator device pumped by a single wavelength, and the capability of tuning, via electric current, the wavelength of the generated pairs. This demonstration is a step towards realization of compact, electrically-tunable wavelength-division multiplexed quantum light generating devices made using CMOS-compatible fabrication and silicon photonics. 

The authors are grateful to Xianshu Luo, Tsung-Yang Liow, Guo-Qiang Patrick Lo (IME, Singapore), Nick Bertone (Optoelectronic Components), and the National Science Foundation for support (ECCS 092539, 1028553, 1153716, 1201308). J. R. O. acknowledges support from A*STAR, Singapore.

\newpage

\newpage
\newpage
\begin{table*}[h]
\begin{minipage}{\textwidth}
\captionof{table}{Simultaneous pair generation from three bands as indicated in Fig. \ref{trans} at room temperature (296.35~K), using a fixed pump wavelength at 1562.6 nm with 200 $\mu$W average power.}\label{table1}
\vspace{0.4cm}
\begin{tabular}{c c c c}
\hline
Band&CAR\footnote{Coincidence-to-accidental ratio. Uncertainties in CAR come from fluctuations in the measured coincidence and accidental counts, and represent one standard deviation values.}
&PGR\footnote{Pair generation rate at the output of waveguide.}(/s)&PGR(/pump pulse)\vspace{1.0 mm}\\
\hline\\
Band 1 & 55$\pm$11 & 6.75$\times$10$^3$ & 1.4$\times$10$^{-3}$ \\
Band 2 & 80$\pm$24 & 1.64$\times$10$^4$ & 3.3$\times$10$^{-3}$ \\
Band 3 & 65$\pm$16 & 1.13$\times$10$^4$ & 2.3$\times$10$^{-3}$ \\
\hline
\end{tabular}
\end{minipage}
\end{table*}
\begin{table*}[h]
\begin{minipage}{\textwidth}
\captionof{table}{Coincidence measurements of pair generation from Band 2 (see Fig.~\ref{trans}) at different temperatures; wavelength shifts with temperature are described in Fig.~\ref{tunecar}.}\label{table2}
\vspace{0.4cm}
\begin{tabular}{c c c c c c c}
\hline
Temp.(K)\footnote{Temperatures were estimated using a measurement of resistance and a thermistor equation: $R_{T} = R_{T0} \textrm{exp} [\frac{\beta (T_{0}-T)}{TT_{0}}]$, where $R_{T}$ is the resistance at a temperature \emph{T}, $\beta=3901$ was the specified material constant of the thermistor, and $R_{T_{0}}=10 \ \mathrm{k\Omega}$ was the resistance at $T_{0}$=298.15~K.}
&CAR\footnote{Coincidence-to-accidental ratio.}
&PGR\footnote{Pair generation rate at the output of the 11-ring section.}(/s)
&PGR(/pump pulse)\vspace{1.0 mm}\\
\hline\\
287.85 &65$\pm$22 &1.27$\times$10$^4$&2.5$\times$10$^{-3}$\\
296.35 &80$\pm$24 &1.64$\times$10$^4$&3.3$\times$10$^{-3}$\\
303.35 &85$\pm$16 &1.12$\times$10$^4$&2.2$\times$10$^{-3}$\\
307.55 &79$\pm$17 &8.59$\times$10$^3$&1.7$\times$10$^{-3}$\\
313.85 &54$\pm$10 &1.16$\times$10$^4$&2.3$\times$10$^{-3}$\\
317.75 &65$\pm$08 &1.44$\times$10$^4$&2.9$\times$10$^{-3}$\\
320.65 &64$\pm$15 &1.39$\times$10$^4$&2.8$\times$10$^{-3}$\\
324.15 &83$\pm$24 &1.17$\times$10$^4$&2.3$\times$10$^{-3}$\\
328.35 &53$\pm$12 &1.23$\times$10$^4$&2.5$\times$10$^{-3}$\\
\hline
\end{tabular}
\end{minipage}
\end{table*}


\begin{thebibliography}{99}
\bibitem{sharping06} J. E. Sharping, K. F. Lee, M. A. Foster, A. C. Turner, B. S. Schmidt, M. Lipson, A. L. Gaeta, and P. Kumar, Opt. Express \textbf{14}, 12388 (2006).
%
\bibitem{harada08} K.-i. Harada, H. Takesue, H. Fukuda, T. Tsuchizawa, T. Watanabe, K. Yamada, Y. Tokura, and S.-i. Itabashi, Opt. Express \textbf{16}, 20368 (2008).
%
\bibitem{melloni11} F. Morichetti, A. Canciamilla, C. Ferrari, A. Samarelli, M. Sorel, and A. Melloni, Nature Commun. \textbf{2}, 296 (2011).
%
\bibitem{ong11} J. R. Ong, M. L. Cooper, G. Gupta, W. M. J. Green, S. Assefa, F. Xia, and S. Mookherjea, Opt. Lett. \textbf{36}, 2964 (2011).
%
\bibitem{mastuda11} N. Mastuda, T. Kato, K-I. Harada, H. Takesue, E. Kuramochi, H. Taniyama, and M. Notomi, Opt. Express \textbf{19}, 19861 (2011).
%
\bibitem{davanco12} M. Davanco, J. R. Ong, A. B. Shehata, A. Tosi, I. Agha, S. Assefa, F. Xia, W. M. J. Green, S. Mookherjea,  and K. Srinivasan, Appl. Phys. Lett. \textbf{100}, 261104 (2011).
%
\bibitem{clemmen09} S. Clemmen, K. Phan Huy, W. Bogaerts, R. G. Baets, Ph. Emplit, and S. Massar, Opt. Express \textbf{17}, 16558 (2009).
%
\bibitem{engin12} E. Engin, D. Bonneau, C. M. Natarajan, A. Clark, M. G. Tanner, R. H. Hadfield, S. N. Dorenbos, V. Zwiller, K. Ohira, N. Suzuki, H. Yoshida, N. Iizuka, M. Ezaki, J. L. O'Brien, and M. G. Thompson, arXiv:1204.4922v2 (2012).
%
\bibitem{jiang12} W. C. Jiang, X. Lu, J. Zhang, O. Painter, and Q. Lin, Frontiers in Optics post-deadline paper FW6C (Rochestor, New York)(2012).
%
\bibitem{azzini12} S. Azzini, D. Grassani, M. J. Strain, M. Sorel, L. G. Helt, J. E. Sipe, M. Liscidini, Matteo Galli, and D. Bajoni, Opt. Express \textbf{20}, 23100 (2012).
%
\bibitem{ong13} J. R. Ong and S. Mookherjea, Opt. Express \textbf{21}, 5171 (2013)

\end{thebibliography}
\end{document}